\documentclass[showpacs]{revtex4}
\usepackage{amssymb,epsfig,graphics}
\begin{document}
\date{\today}
\title{Thermo--inertial bouncing of a relativistic collapsing sphere: A
numerical model}

\author{L.~Herrera}
\email{laherrera@telcel.net.ve}
\altaffiliation[Postal Address: ]{Apartado 80793, Caracas 1080A, Venezuela}
\author{A. Di Prisco}
\email{adiprisc@fisica.ciens.ucv.ve}
\affiliation{Centro de F\'\i sica Te\'orica y Computacional.
Facultad de Ciencias.
Universidad Central de Venezuela.
Caracas. Venezuela.}

\author{W.~Barreto}
\email{wbarreto@ula.ve}
\affiliation{Centro de F\'\i sica Fundamental,
Facultad de Ciencias, Universidad de los Andes, M\'erida, Venezuela.}

\begin{abstract}
We present a numerical model of a collapsing radiating sphere, whose
boundary surface undergoes  bouncing due to a  decreasing of its inertial
mass density (and, as expected from the
equivalence principle, also of the ``gravitational'' force term)   produced
by the ``inertial'' term of the transport
equation. This model exhibits for the first time  the consequences of  such
an effect, and shows that under physically reasonable conditions this
decreasing of the gravitational term
in the dynamic equation may be large enough as to revert the collapse and
produce a  bouncing of the boundary surface of the sphere.
\end{abstract}
\pacs{04.40.-b; 04.40.Dg; 95.30.Lz; 97.60.-s.}
\maketitle
\section{Introduction}
In the study of gravitational collapse of massive stars, the inclusion of
dissipative processes (in particular neutrino emission) is enforced by the
fact that they provide the only
plausible mechanism to carry away the bulk of binding energy, leading to a
neutron star or black hole \cite{1}. On the other hand, in cores of densities
about $10^{12}g\,\,cm^{-3}$ the mean free
path of neutrinos becomes small enough as to justify the use of diffusion
approximation \cite{3,4}. This seems to be confirmed by the observational
data collected from supernova 1987A,
which indicates that the radiation transport regime prevailing during
the emission process, is closer to the diffusion approximation than to the
streaming out limit \cite{5}.

Motivated by the comments above, in a recent paper \cite{Herrera1}, the
Misner and Sharp approach to the study of adiabatic gravitational collapse
\cite{MisnerSharp} was extended as to
include dissipation in, both, the streaming out and diffusion approximation
(for the case of pure free streaming approximation see \cite{Misner}). Then
from the coupling of the
dynamical equation to a causal transport equation in the context of
M\"uller--Israel--Stewart theory  \cite{Muller67,IsSt76} it was obtained that
the effective inertial mass density of a
fluid element and the gravitational force term in the dynamical equation,
reduce by a factor which depends on dissipative variables. This reduction,
in its turn, might lead to the
bouncing of the collapsing sphere, as discussed in \cite{Herrera1}.

As can be seen from inspection of the transport equation, such an effect is
directly related to the presence of the inertial term $Ta_{\beta}$ in the
transport equation.  This explains why
we refer to  such a bouncing as ``thermo--inertial''.

It is our purpose in this work to present a numerical model of a radiating
collapsing sphere, where the above mentioned effect produces the bouncing
of the boundary surface of
the sphere, for physically acceptable values of all variables.

 Since we are mainly concerned with time scales of the order of magnitude
of (or even smaller than) the hydrostatic time scale, as in the quick
collapse phase  preceding neutron star formation, we cannot rely on the
quasistatic approximation, and therefore the full dynamic description has
to be used \cite{8A,9A}. This
implies that we have to  appeal to a hyperbolic theory of
dissipation.								
The use of a hyperbolic theory of dissipation is further justified  by the
necessity of overcoming the difficulties inherent to parabolic theories
(see references \cite{6}--\cite{22} and references therein).

The plan of the paper is as follows. In the next section we define the
conventions and present the dynamical equation  coupled to the transport
equation. The model to be considered
as well as the strategy for the numerical integration is presented in
Section III. Finally, a discussion of results is presented in Section IV.

\section{The dynamical equation of the dissipative fluid}
We consider a spherically symmetric distribution of collapsing
fluid (for simplicity we shall consider the pressure to be locally isotropic)
undergoing dissipation in the form of heat flow, bounded by a
spherical surface $\Sigma$.
We assume the interior metric to $\Sigma$ to be comoving, shear free for
simplicity, and spherically symmetric, accordingly it may be written as
\begin{equation}
ds^2=-A^2(t,r)dt^2+B^2(t,r)(dr^2+r^2d\theta^2+r^2\sin^2\theta d\phi^2),
\label{3}
\end{equation}
and hence we have for the four velocity $V^{\alpha}$ and the heat flux
vector $q^{\alpha}$
\begin{equation}
V^{\alpha}=A^{-1}\delta^{\alpha}_0, \;\; q^{\alpha}=q\delta^{\alpha}_1 \;\;
. \label{4}
\end{equation}

Then it can be shown \cite{Herrera1} that the following equation can be
found from Bianchi identities
\begin{eqnarray}
(\mu+P)D_tU=-(\mu+P)\left[m+4\pi P R^3 \right]\frac{1}{R^2}
-E^2 D_R P 
-E\left[5qB\frac{U}{R}+BD_tq\right],
\label{29}
\end{eqnarray}
where $\mu$ is the energy density, $P$ the pressure,
\begin{equation}
D_t=\frac{1}{A}\frac{\partial}{\partial t} \label{16},
\end{equation}
the proper radial derivative $D_R$,
constructed from the radius of a spherical surface, as measured from its
perimeter inside $\Sigma$, being
\begin{equation}
D_R=\frac{1}{R^{\prime}}\frac{\partial}{\partial r}, \label{23a}
\end{equation}
with
\begin{equation}
R=rB, \label{23aa}
\end{equation}
and where dots and primes denote derivatives with respect to $t$ and $r$
respectively.
The velocity $U$ of the collapsing fluid is defined as
\begin{equation}
U=rD_tB<0 \;\; (in \; the \;  case \; of \; collapse). \label{19}
\end{equation}
Also, the mass function $m(t,r)$ of Cahill and McVittie \cite{Cahill} is
obtained from the Riemann tensor component ${R_{23}}^{23}$ and is for
metric (\ref{3})
\begin{equation}
m(t,r)=\frac{(rB)^3}{2}{R_{23}}^{23}=\frac{r^3}{2}\frac{B\dot{B}^2}{A^2}
-\frac{r^3}{2}
\frac{B^{\prime 2}}{B}-r^2B^{\prime}, \label{8a}
\end{equation}
$E$ is defined as
\begin{equation}
E=\frac{(rB)^{\prime}}{B}=\left[1+U^2-\frac{2m(t,r)}{rB}\right]^{1/2}.
\label{20}
\end{equation}

Next, the corresponding  transport equation for the heat flux reads \cite
{Muller67,IsSt76}
\begin{equation}
\tau
h^{\alpha\beta}V^{\gamma}q_{\beta;\gamma}+q^{\alpha}=-\kappa h^{\alpha\beta}
(T_{,\beta}+Ta_{\beta}) -\frac 12\kappa T^2\left( \frac{\tau
V^\beta }{\kappa T^2}\right) _{;\beta }q^\alpha ,  \label{21}
\end{equation}
where $h^{\mu \nu }$ is the projector onto the three space orthogonal to $%
V^\mu $,
$\kappa $  denotes the thermal conductivity, and  $T$ and  $\tau$
denote temperature and relaxation time
respectively.
Observe that due to the symmetry of the problem, equation (\ref{21}) only
has one independent component, which may be written as:
\begin{eqnarray}
BD_tq=-\frac{\kappa T}{\tau E}D_tU-\frac{\kappa T^{\prime}}{\tau B}-
\frac{qB}{\tau}(1+\frac{\tau U}{R})
-\frac{\kappa
T}{\tau E}\left[m+4\pi P\right]R^{-2}-
\frac{\kappa T^2 q B}{2A\tau}\left(\frac{\tau}{\kappa T^2}\right)\dot{}
-\frac{3U Bq}{2R}, \label{V3}
\end{eqnarray}

Then coupling (\ref{29}) to (\ref{V3}) one obtains (some misprints in
eq.(39) in \cite{Herrera1} has been corrected here)
 \begin{eqnarray}
(\mu+P)(1-\alpha)D_tU=F_{grav}(1-\alpha)+F_{hyd}
+\frac{E\kappa T^{\prime}}{\tau B}+\frac{EqB}{\tau}-\frac{5qBE U}{2R}
+\frac{\kappa ET^2 q B}{2A \tau}\left(\frac{\tau}{\kappa T^2}\right)\dot{},
\label{V4}
\end{eqnarray}
where $F_{grav}$ and $F_{hyd}$ are defined by
\begin{equation}
F_{grav}=-(\mu+P)\left[m+4\pi P R^3\right]\frac{1}{R^2}, \nonumber \\
\label{grav}
\end{equation}
and
\begin{equation}
F_{hyd}=
-E^2 D_R P,
\label{hyd}
\end{equation}
with $\alpha$ given by
\begin{equation}
\alpha=\frac{\kappa T}{\tau (\mu+P)}.
\label{alpha}
\end{equation}
Thus as $\alpha$ tends to $1$, the effective inertial mass density of the
fluid element tends to zero.
Furthermore observe that $F_{grav}$ is also multiplied by the factor
$(1-\alpha)$. Indicating that the effective
gravitational attraction on any fluid element decreases by the same factor
as the effective inertial mass (density). This of
course is to be expected, from the equivalence principle. It is also worth
mentioning that $F_{hyd}$ is in principle independent
(at least explicitly) on this factor.

With these last comments in mind, let us now imagine the following
situation. As far as the right hand side of (\ref{V4}) is negative, the
system
keeps collapsing. However, let us assume that the  collapsing sphere
evolves in such a way that, for some region of the sphere, the value of
$\alpha$ increases  and  approaches the
critical value of
$1$. Then, as this process goes on, the ensuing decreasing of the
gravitational force term would eventually lead to a change of the sign
of the right hand side of (\ref{V4}). Since that would happen for small
values of the
effective inertial mass density, that would imply a strong bouncing of that
part of the  sphere, even for a small absolute value of the
right hand side of (\ref{V4}).

In the next section a model will be presented where the effect above appears
explicitly. For simplicity we shall consider a particular case of the
transport equation, corresponding to the
so called truncated  version, in which case the last term on the right of
(\ref{21}) is absent \cite{8,19}. In this case, (\ref{V4})
becomes
 \begin{eqnarray}
(\mu+P)(1-\alpha)D_tU=F_{grav}(1-\alpha)+F_{hyd}
+\frac{E\kappa T^{\prime}}{\tau B}+\frac{EqB}{\tau}-\frac{4qBE U}{R}.
\label{V41}
\end{eqnarray}

\section{The model}
In this section we shall present a numerical model where the decreasing of
the effective mass mentioned in the previous section will produce a
bouncing during the evolution of a
dissipative sphere. For simplicity we shall assume our fluid to be
shear--free and conformally flat, and also that a relevant increase of
$\alpha$ takes place only at the
boundary surface of the sphere. Thus we shall need only to integrate at the
boundary surface, implying that we shall deal with ordinary differential
equations for variables defined on
that surface.
\subsection{The general form of the metric and the field equations}
 If the fluid sphere is shear--free and  conformally flat, the
metric functions take the form \cite{CF}
\begin{equation}
A=\left[C_1(t)r^2+1\right]B
\label{A}
\end{equation}
and
\begin{equation}
B=\frac{1}{C_2(t)r^2+C_3(t)},
\label{B}
\end{equation}
where $C_1$, $C_2$ and $C_3$ are arbitrary functions of $t$. Although
the shear free and the conformally flat conditions are introduced
here in order to simplify calculations, it is worth noticing that these
conditions generalize physical assumptions widely used in astrophysics.
Indeed, the shear free condition in the Newtonian regime describes the
homologous evolution and has been extensively considered in general
relativity \cite{SFCF}. On the other hand it is well known that the
conformally flat condition implies in the perfect fluid
case the homogeneity of the energy density distribution.

For the numerical integration we shall need to write all variables in
dimensionless form, accordingly we shall redefine the metric functions
$C_1$ and $C_2$ by ($C_3$ is already
dimensionless):

\begin{equation}
C_{1,2}\rightarrow\frac{C_{1,2}}{r_{\Sigma}^2},
\label{dimens}
\end{equation}
where $r=r_{\Sigma}=constant$ defines the boundary surface of the fluid sphere.

In terms of these dimensionless functions, $A$ and $B$ become
\begin{equation}
A=\left[C_1(t)(r/r_{\Sigma})^2+1\right]B
\label{A1}
\end{equation}
and
\begin{equation}
B=\frac{1}{C_2(t)(r/r_{\Sigma})^2+C_3(t)}.
\label{B1}
\end{equation}

Then the following expressions for the physical variables are obtained from
Einstein equations
\begin{equation}
\mu r_{\Sigma}^2=\frac{3}{8 \pi}\left(\frac{\dot{C}_2
(r/r_{\Sigma})^2+\dot{C}_3}{C_1(r/r_{\Sigma})^2+1}\right)^2+
\frac{3}{2 \pi}C_2C_3,
\label{3n}
\end{equation}
\begin{eqnarray}
r_{\Sigma}^2
P=\frac{1}{8\pi(C_1\left(r/r_{\Sigma}\right)^2+1)^2}[ 2(\ddot{C}
_2(r/r_{\Sigma})^2+\ddot{C}_3)(C_2(r/r_{\Sigma})^2+C_3)\nonumber \\
-3(\dot{C}_2(r/r_{\Sigma})^2+\dot{C}_3)^2
-2\frac{\dot{C}_1(r/r_{\Sigma})^2}{C_1(r/r_{\Sigma})^2+1}
(\dot{C}_2(r/r_{\Sigma})^2+\dot{C}_3)\nonumber\\
\left(C_2(r/r_{\Sigma})^2+C_3 \right)]
+\frac{1}{2\pi(C_1(r/r_{\Sigma})^2+1)}\left[C_2(C_2-2C_1C_3)(r
/r_{\Sigma})^2\right. \nonumber \\
\left. +C_3(C_1C_3-2C_2)\right],
\label{4n}
\end{eqnarray}
\begin{equation}
 q r_{\Sigma}^2=\frac{1}{2\pi}(r/r_{\Sigma})(\dot{C}_3C_1-\dot{C}_2)
\left(\frac{C_2(r/r_{\Sigma})^2+C_3}{C_1(r/r_{\Sigma})^2+1}\right)^2, \label{5n}
\end{equation}
where from now on  dot denotes
derivative with respect to $t/r_{\Sigma}$.
\subsection{The surface equations}

Next, from (\ref{23aa}) we obtain for the dimensionless proper radius of
the sphere
\begin{equation}
R_{\Sigma}=\frac{1}{C_{2}+C_{3}},
\label{1B}
\end{equation}
and from (\ref{20}) evaluated at the boundary surface
\begin{equation}
E_{\Sigma}=\frac{C_{3}-C_{2}}{C_{3}+C_{2}}.
\label{2B}
\end{equation}
Solving these two equations we obtain
\begin{equation}
C_{2}=\frac{1-E_{\Sigma}}{2R_{\Sigma}}
\label{3B}
\end{equation}
and
\begin{equation}
C_{3}=\frac{1+E_{\Sigma}}{2R_{\Sigma}}.
\label{4b}
\end{equation}
On the other hand we have
\begin{equation}
A_{\Sigma}=(C_{1}+1)R_{\Sigma},
\label{4B}
\end{equation}
and  from (\ref{19})
\begin{equation}
U_{\Sigma}=-\frac{\dot C_{2}+\dot C_{3}}{(C_{1}+1)(C_{2}+C_{3})},
\label{5B}
\end{equation}
where, again, dots denote derivatives with respect to the dimensionless
time $t/r_{\Sigma}$.
Using (\ref{1B}) and (\ref{4B}) in (\ref{5B}) we may write
\begin{equation}
\dot R_{\Sigma}=A_{\Sigma}U_{\Sigma},
\label{first}
\end{equation}
which is our first equation at the surface.

Next, we use the total loss of mass
equation which can be easily derived from (\ref{8a}) and the junction condition
\begin{equation}
P_{\Sigma}=(qB)_{\Sigma},
\label{jc}
\end{equation}
to obtain (see \cite{Herrera1}) for details)
\begin{equation}
\dot M_{\Sigma}=-Q(t)A_{\Sigma}R_{\Sigma}(U_{\Sigma}+E_{\Sigma}),
\label{second}
\end{equation}
where $M_{\Sigma}$ is the dimensionless mass, $Q(t)\equiv 4\pi
q_{\Sigma}R_{\Sigma}^{2}$ and $q_{\Sigma}$ denotes the dimensionless heat
flow  $q r_{\Sigma}^2$, evaluated at
the boundary surface. This is our second surface equation.

Finally, in order to obtain the third surface equation we proceed as
follows. From the equations (\ref{4n}) and (\ref{jc}) in \cite{CF}, it can be shown
that
\begin{equation}
\dot U_{\Sigma}=\frac{1}{2R_{\Sigma}}
\left[
A_{\Sigma}\left(3E^{2}_{\Sigma}-1-U^{2}_{\Sigma}-2R_{\Sigma}Q(t)\right)
+2E_{\Sigma}\left(A_{\Sigma}-2R_{\Sigma}\right)
\right].
\label{third}
\end{equation}
This is the third equation to be integrated at the surface of the distribution.

Thus we have a system of three equations (\ref{first}), (\ref{second}) and (\ref{third}) for the
five unknown functions of time $R_{\Sigma}, A_{\Sigma}, U_{\Sigma},
E_{\Sigma}$ and $Q$. In order to
integrate such a system, we shall prescribe the ``luminosity'' ($Q$),  and
obtain a constraint equation from (\ref{V41}), on what we shall elaborate
as follows .

From the dynamic equation (\ref{V41}), using the boundary condition (\ref{jc})
 we obtain the pressure gradient
at the surface
\begin{equation}
P^{'}_{\Sigma}=-\frac{R_{\Sigma}}{E_{\Sigma}}
\left\{
\left[1-\alpha_{\Sigma}\right]\left[\mu_{\Sigma}+P_{\Sigma}\right]
[4\pi R_{\Sigma}(\mu_{\Sigma}/3+P_{\Sigma})+\dot U_{\Sigma}/A_{\Sigma}]
-{\mathcal{T}_{\Sigma}}
\right\},
\label{grad}
\end{equation}
where primes and dots denote derivatives with respect to the dimensionless
variables $r/r_{\Sigma}$ and $t/r_{\Sigma}$ respectively,
\begin{equation}
\mu_{\Sigma}=\frac{3M_{\Sigma}}{4\pi R_{\Sigma}^{3}},
\end{equation}
\begin{equation}
P_{\Sigma}=\frac{Q(t)}{4\pi R_{\Sigma}},
\end{equation}
(observe that  $\mu_{\Sigma}$ and $P_{\Sigma}$ denote the dimensionless
expressions for the energy density and pressure evaluated at the boundary
surface, i.e. these variables
multiplied by $r_{\Sigma}^2$) and
\begin{equation}
{\mathcal{T}_{\Sigma}}=\alpha_{\Sigma}\left(\mu_{\Sigma}^{'}+P_{\Sigma}^{'}
\right)\frac{E_{\Sigma}}{R_{\Sigma}}+
\frac{E_{\Sigma}Q(t)}{4\pi
R_{\Sigma}\tau}-\frac{4QE_{\Sigma}U_{\Sigma}}{4\pi R_{\Sigma}^{2}},
\label{B7}
\end{equation}
where we have used $kT=\alpha \tau(\mu+P)$ (conveniently adimensionalised)
and have assumed for simplicity that $\alpha^{\prime}_{\Sigma}=0$.
Finally, using the field equations (\ref{3n})--(\ref{5n}) we obtain
the following expression for $A_\Sigma$
\begin{equation}
A_\Sigma=\frac{\tau R_{\Sigma}[4\alpha_{\Sigma}(3M_\Sigma/R_\Sigma+QR_\Sigma)
-\dot QR_\Sigma(1+\alpha_{\Sigma})]}
{2\tau\alpha_{\Sigma}(1+E_\Sigma)(QR_\Sigma+3M_\Sigma/R_\Sigma)
-QR_\Sigma(R_\Sigma-
7U_\Sigma\tau)}.
\label{fourth}
\end{equation}
\subsection{Strategy of integration}
 The integration scheme is now an easy shot:
Giving initial conditions for $R_{\Sigma}$, $U_{\Sigma}$ and $M_{\Sigma}$,  and
prescribing $\alpha_{\Sigma}$ and $Q(t)$,
we can integrate numerically, equations (\ref{first}), (\ref{second}),
(\ref{third}),
with  the constraint equation (\ref{fourth}).

The form of $\alpha_{\Sigma}$ is suggested by the very idea underlying the
motivation of this work, namely: the fact that as $\alpha$ increases, the
ensuing reduction of the
gravitational term in the dynamical equation may lead to a bouncing of the
sphere. Accordingly, we shall take for $\alpha_{\Sigma}$ a smooth function
of time, rising from zero to some
value below the critical one ($\alpha=1$).
\subsection{Model}
We have ran  a large number of models exhibiting bouncing, under physically
reasonable conditions, corresponding to a wide range of initial data and
values of the parameters, and very
different choices of $Q(t)$ and $\alpha(t)$. For all these choices,
the qualitative behaviour associated to the increase of $\alpha$ is
essentially the same. From them we have selected the
following model.

The initial conditions are
$$R_{\Sigma}(0)=20,$$
$$M_{\Sigma}(0)=1,$$
$$U_{\Sigma}(0)=-0.1,$$
with $\tau=0.1$. These values  correspond to a sphere with an initial
radius of the order of $400$ Km, an initial mass of the order of 10 solar
masses and a relaxation time of the
order of
$10^{-4}$ seconds.

The sphere is assumed to be radiating according to
$$Q(t)=Q_0e^{-(t-t_m)^2/\sigma},$$
where $Q_0=0.001$, $t_m=0.5$ and  $\sigma=0.005$, producing a total mass
ejection of the order of $0.1$\%.

For $\alpha$ we choose
$$\alpha(t)=\alpha_m/(e^{-(t-t_m)/\sigma}+1),$$
 with  $\alpha_m$ ranging from $0$ to $1$.
\section{Discussion}
The influence of  pre--relaxation effects on gravitational collapse has
been brought out in many works in last decade \cite{pre}, however the
specific effect of bouncing, associated
with the decreasing of the effective inertial mass density, produced by the
increasing of $\alpha$, had not been illustrated until now. It is worth
stressing that $\alpha$--terms in Eq. (\ref{V4}) come from the inertial
factor $Ta_\beta$ in Eq. (\ref{21}).

\begin{figure}
\centerline{\epsfxsize=4.5in\epsfbox{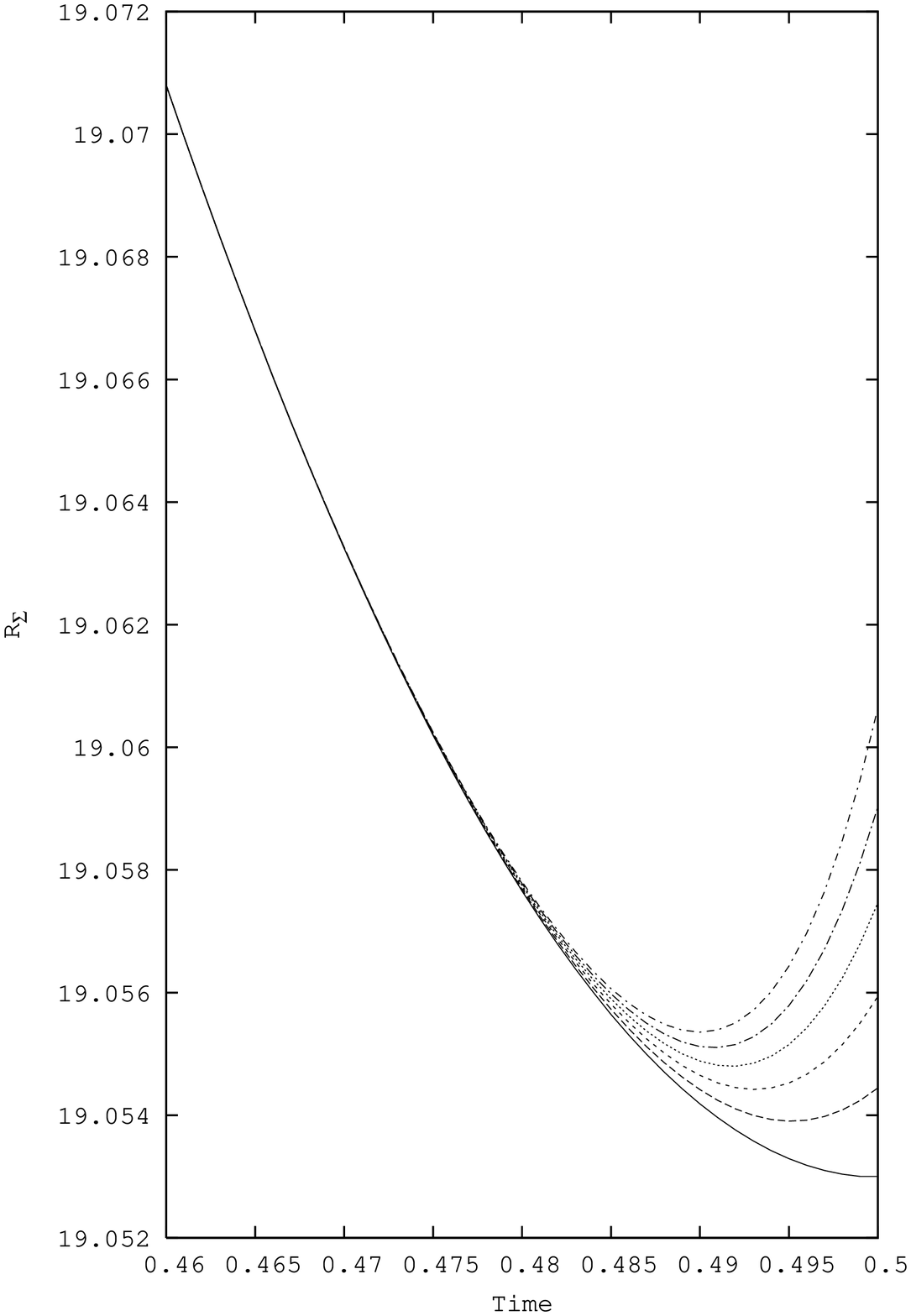}}
\caption{Evolution of $R_{\Sigma}$ for different values of $\alpha_m$:
0.0 (solid line); 0.2 (large dashed line); 0.4 (short dashed line);
0.6 (dotted line); 0.8 (dot--large dashed line)  and 1 (dot--short dashed
line).}

\label{fig:r}
\end{figure}

\begin{figure}
\centerline{\epsfxsize=4.5in\epsfbox{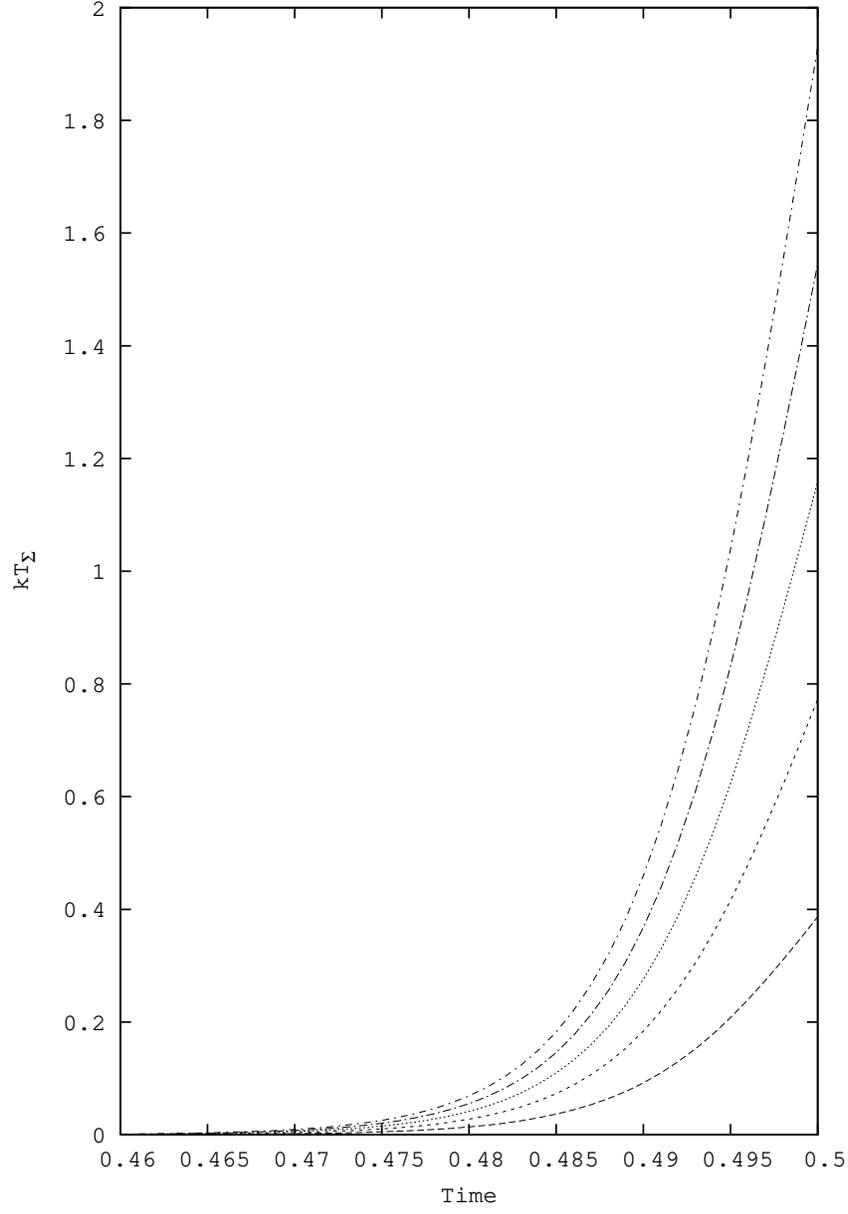}}
\caption{Temperature at the surface (multiplied by $\kappa 10^{7}$) evolution
for different values of $\alpha_m$:
0.0 (solid line); 0.2 (large dashed line); 0.4 (short dashed line);
0.6 (dotted line); 0.8 (dot--large dashed line)  and 1 (dot--short dashed
line).}
\label{fig:t}
\end{figure}

In this work we provide a numerical model of such bouncing, by assuming an
increasing of $\alpha$ at the boundary surface. We have concentrated
the increase of $\alpha$ on the boundary surface to illustrate
the effect, the remaining of the sphere is assumed to be dissipating
at much lower values of $\alpha$. Of course, the increasing of $\alpha$
may in principle occur at any region of the sphere and even in more
that one, simultaneously. The results of our
integration is deployed in the figures {\ref{fig:r}}--{\ref{fig:t}}, which
exhibit the evolution of different variables with respect to the
dimensionless time $t/r_{\Sigma}$.

\begin{figure}
\centerline{\epsfxsize=4.5in\epsfbox{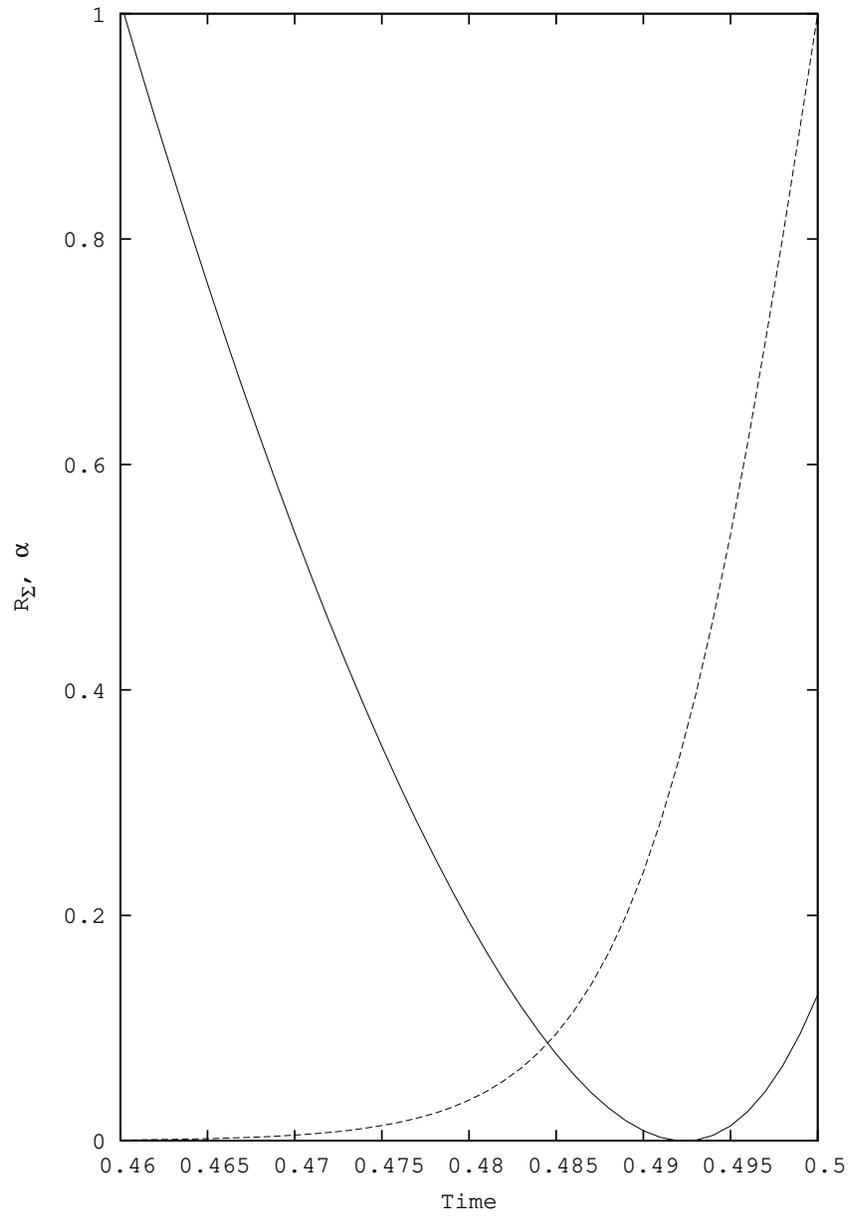}}
\caption{Evolution of $R_{\Sigma}$ (continuous line) and $\alpha$ (dashed line)
for $\alpha_m=0.5$. The curves were normalized (and
shifted only for $R_\Sigma$)
in order to display them together.}
\label{fig:Ralpha}
\end{figure}

 Figure  {\ref{fig:r}}   shows the evolution of $R_{\Sigma}$ for different
values of $\alpha_m$ from $0$ to $1$, the bouncing is clearly exhibited
as well as its
dependence on $\alpha$. Figure {\ref{fig:Ralpha}} emphasizes further the
link between the increasing of $\alpha$ and the bouncing.

\begin{figure}
\centerline{\epsfxsize=4.5in\epsfbox{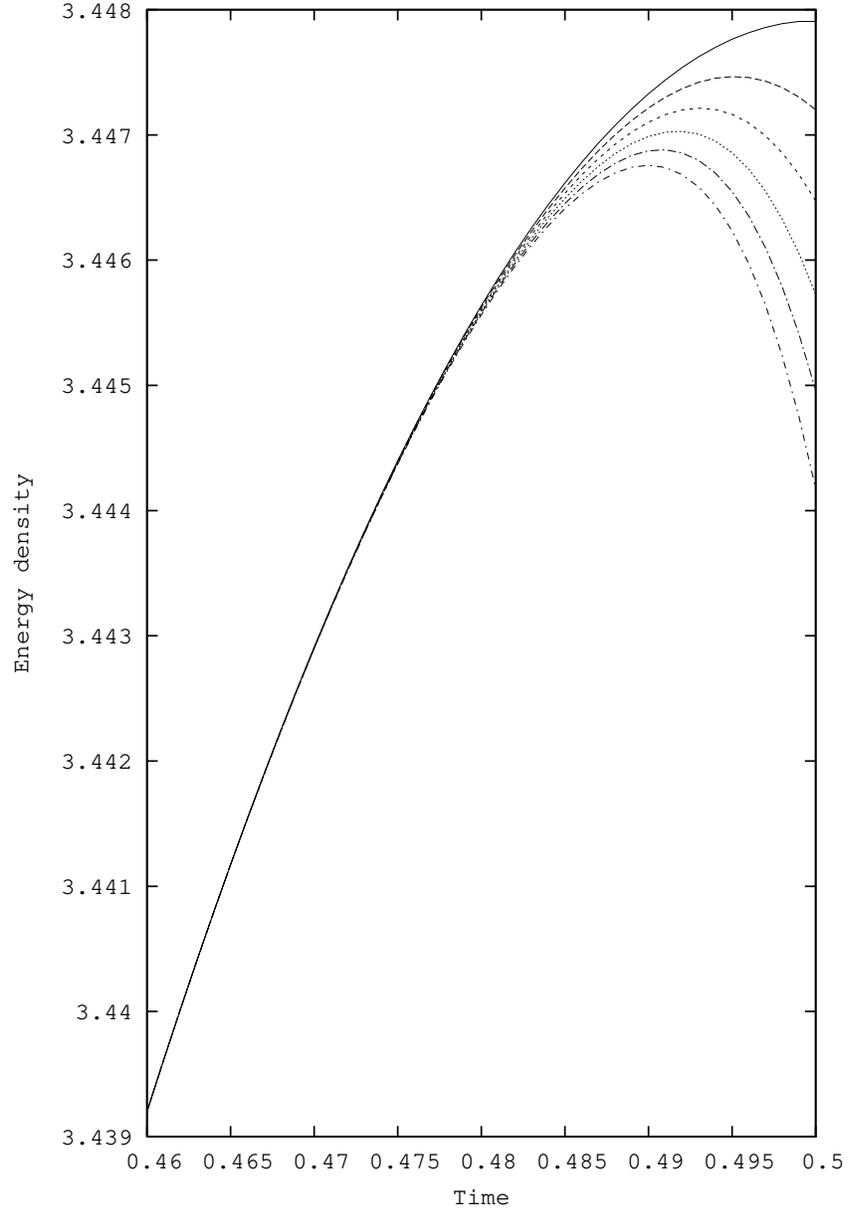}}
\caption{Energy density at the surface (multiplied by $10^{5}$)
evolution for different values of $\alpha_m$:
0.0 (solid line); 0.2 (large dashed line); 0.4 (short dashed line);
0.6 (dotted line); 0.8 (dot--large dashed line)  and 1 (dot--short dashed
line).}
\label{fig:d}
\end{figure}

Figures {\ref{fig:d}}--{\ref{fig:t}}, shows the  behaviour
of (dimensionless) energy density, pressure, heat flow and temperature,
evaluated at the
boundary surface. Their values are always regular and satisfy the physical
conditions $\rho>P>0$.

The dimensionless quantity $\kappa T_{\Sigma}$ plotted in
Figure {\ref{fig:t}}  is, in conventional units,
\begin{equation}
\kappa T_{\Sigma}=2\,\, 10^6 \frac{G}{c^5}[\kappa][T_{\Sigma}]
\label{temp1}
\end{equation}
 with $G$ and $c$ denoting the gravitational constant and the speed of
light, and where $[\kappa]$ and $[T]$ denote the numerical values of
conductivity and temperature in
$g \,\, cm^{-3}\,\, K^{-1}$ and
$K$ respectively. Therefore the maximum values of
$\kappa T_{\Sigma}$ reached just after the bouncing, correspond to
\begin{equation}
[\kappa][T_{\Sigma}] \approx 10^{46}
\label{temp2}
\end{equation}
which may be obtained with $[T_{\Sigma}]\approx 10^{12}$ and $[\kappa]
\approx 10^{34}$. These values are well within the acceptable range for
those variables
in a pre--supernovae event \cite{Ma}.

Thus we have seen that a relatively simple model, whose physical variables
exhibit good behaviour and have acceptable numerical values, may serve to
illustrate
the bouncing of a dissipating self--gravitating
sphere, produced by the decreasing of its effective inertial mass density
associated to an increasing of $\alpha$.

Nevertheless, in spite of the appeal of the presented model, we are well
aware that invoking such an effect to describe a
specific observed phenomena, would
require a much more detailed astrophysical setting. This, however,  is out
of the scope of this paper.

\begin{acknowledgments}
WB was benefited from research support from
FONACIT under grant S1--98003270.
Computer time was provided by the Centro Nacional de
  C\'alculo Cient\'\i fico, Universidad de Los Andes (CeCalcULA).
\end{acknowledgments}

\newpage
\begin{figure}
\centerline{\epsfxsize=4.5in\epsfbox{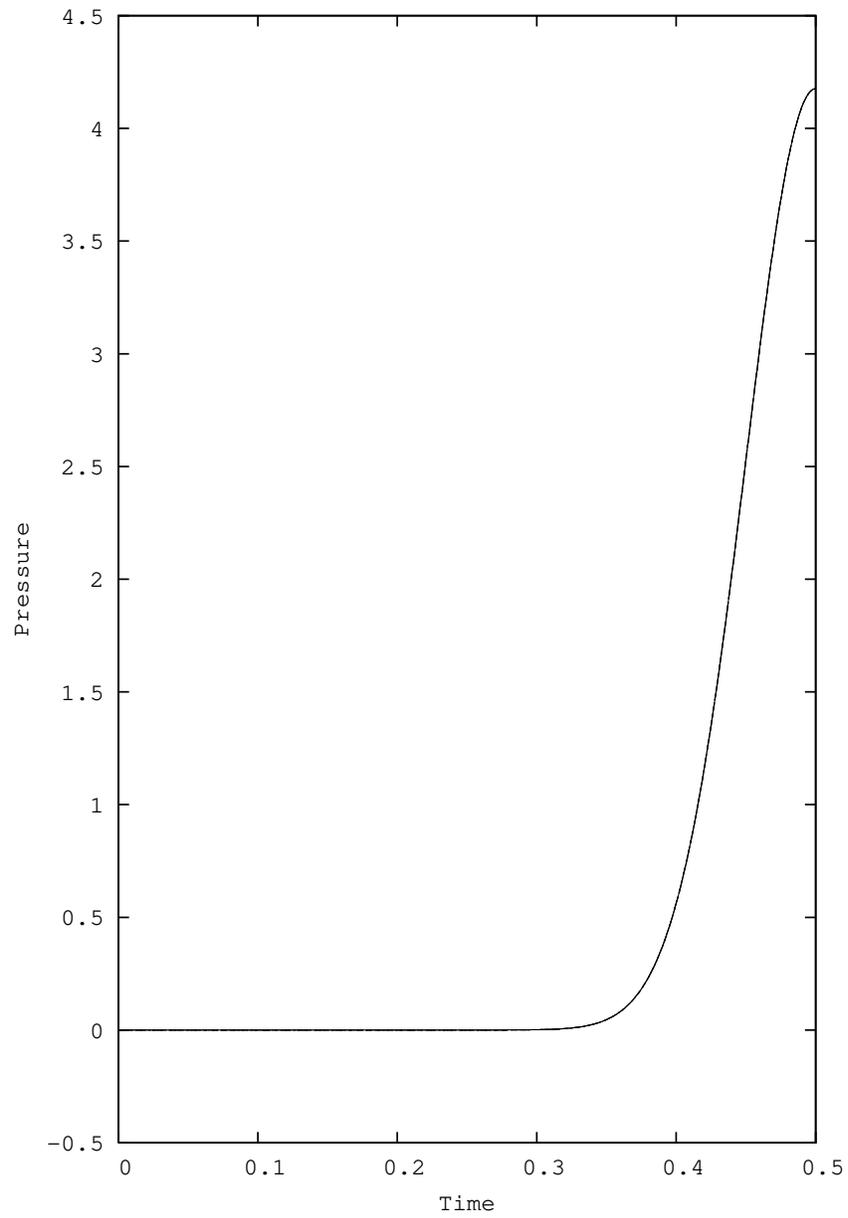}}
\caption{Pressure at the surface (multiplied by $10^{6}$)
 evolution for the same values of $\alpha$ as in previous figure. They all
overlap within the approximation of the plotter.
}
\label{fig:p}
\end{figure}
\begin{figure}
\centerline{\epsfxsize=4.5in\epsfbox{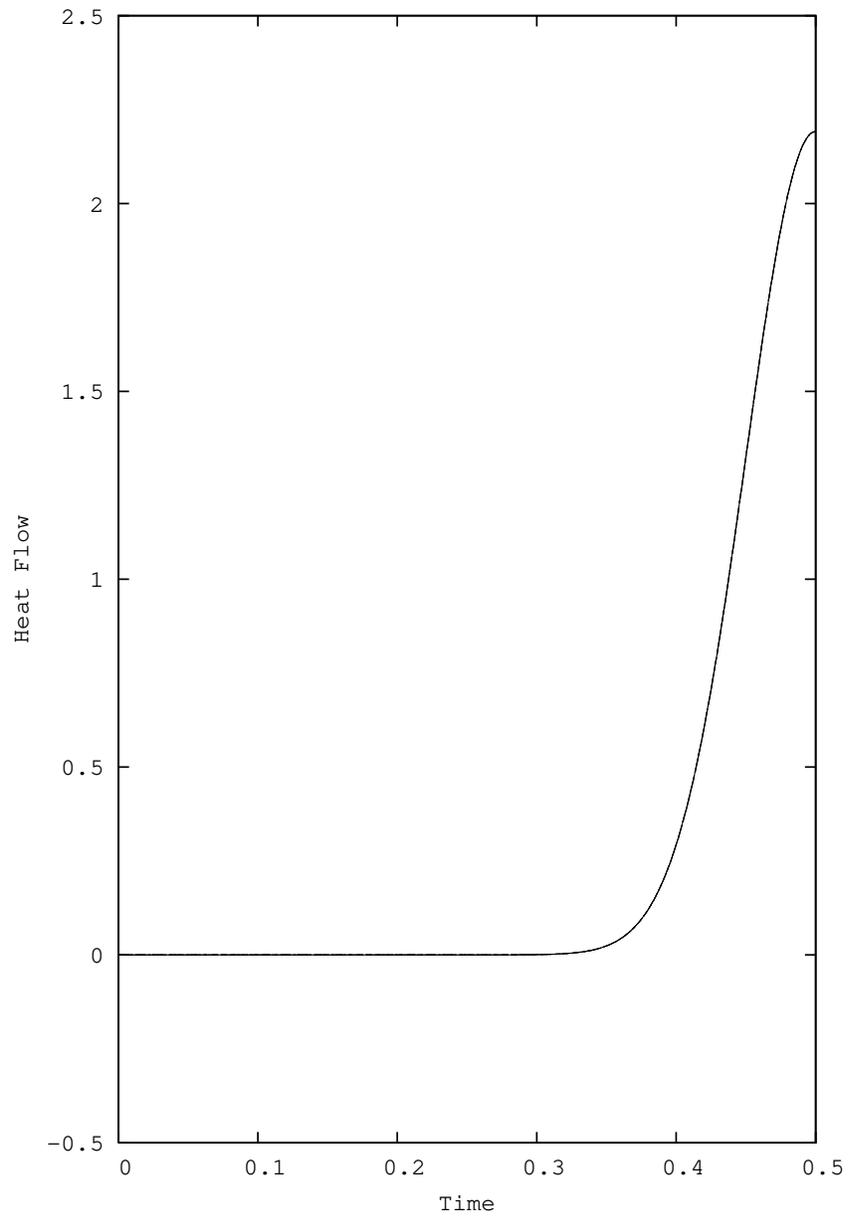}}
\caption{Heat flow at the surface (multiplied by $10^{7}$)
 evolution for the same values of $\alpha_m$. They all overlap within the
approximation of the plotter.
}
\label{fig:q}
\end{figure}

\end{document}